\PassOptionsToPackage{numbers,sort & compress}{natbib}
\documentclass[11pt,showpacs,showkeys]{revtex4}
\usepackage{bbold}
\usepackage{float}
\usepackage[pagebackref=false,colorlinks,linkcolor=blue,citecolor=magenta]{hyperref}
\usepackage{subfigure}
\usepackage{graphicx}
\usepackage{amssymb}
\usepackage{amsmath}
\usepackage{amsfonts}
\usepackage{mathrsfs}
\usepackage{txfonts}
\usepackage{fontenc}
\usepackage{tone}
\DeclareMathAlphabet{\mathcal}{OMS}{zplm}{m}{n}
\usepackage{euscript}
\usepackage{subfigure}
\usepackage{bm} 
\usepackage{graphicx}
\usepackage{color}
\usepackage{multirow}

\begin{document}
	\title{ The Dunkl oscillator on a space of nonconstant curvature: an exactly solvable quantum model with reflections}
\author{Angel Ballesteros $^{1}$}
\email[ ]{angelb@ubu.es}
\author{Amene Najafizade$^{2}$}
\email[ ]{Najafizade1816@gmail.com}
\author{Hossein Panahi$^{2}$}
\email[ ]{t-panahi@guilan.ac.ir}
\author{Hassan Hassanabadi$^{3}$}
\email[ ]{h.hasanabadi@shahroodut.ac.ir}
\author{Shi-Hai Dong $^{4,5}$}
\email[ ]{dongsh2@yahoo.com}
\affiliation{$^{1}$Departamento de F\'{i}sica, Universidad de Burgos, 09001 Burgos, Spain}
\affiliation{$^{2}$Department of Physics, University of Guilan, Rasht 51335-1914, Iran }
\affiliation{$^{3}$Faculty of Physics, Shahrood University of Technology, Shahrood, Iran}
\affiliation{$^{4}$Huzhou University, Huzhou, 313000, P. R. China}
\affiliation{$^{5}$Centro de Investigaci\'{o}n en Computaci\'{o}n, Instituto Polit\'{e}cnico Nacional, UPALM, CDMX 07700, Mexico}

\date{\today}
\begin{abstract}

We introduce the Dunkl-Darboux III oscillator Hamiltonian in N dimensions as a $\lambda-$deformation of the Dunkl oscillator. This deformation is interpreted as the introduction of a non-constant curvature on the underlying space or, equivalently, as a quadratic position-dependent mass for the Dunkl oscillator. This new ND quantum model is shown to be exactly solvable, and its eigenvalues and eigenfunctions are explicitly presented. It is shown that in the 2D case both Darboux III and Dunkl oscillators can be coupled with a constant magnetic field, thus giving rise to two new exactly solvable quantum systems in which the effect of the $\lambda-$deformation and of the Dunkl derivatives on the Landau levels can be studied. Finally, the full 2D Dunkl-Darboux III oscillator is coupled with the magnetic field and shown to define  an exactly solvable Hamiltonian, where the interplay between the $\lambda-$deformation and the magnetic field is explicitly illustrated.

\end{abstract}
\pacs{02. 30. Jr, 03. 65. -w, 03. 65. Ge}
\maketitle



\section{Introduction}

The  isotropic harmonic oscillator on the $N$-dimensional Euclidean space constitutes the paradigm of maximally superintegrable (classical and quantum) Hamiltonian systems, whose superintegrability is deeply connected with the existence of a maximal set of symmetries for the Hamiltonian. As it is well-known, when the oscillator potential (including also centrifugal terms) is suitably defined on N-dimensional spaces with constant curvature, the superintegrability of the system is preserved, together with the dimensionality of the algebra of symmetries (see for instance~\cite{FMSUW65,Ev90b,Ev91,10,20,RS,21,BH07} and references therein).

Interestingly enough, a superintegrable version of the  isotropic oscillator on a N-dimensional (ND) space with nonconstant curvature was found in~\cite{physicaD,PLA} (see also~\cite{cq0,cq1,cq2,cq3,ref15,latiniJPCS,bert1,cq6,cq7} and references therein). This space is the so-called ND Darboux III space, a conformally flat space with metric 
 \begin{equation}\label{eq3}
 \mathrm{d}\mathbf{s}^2=(1+\lambda\, \mathbf{x}^2) \, \mathrm{d}\mathbf{x}^2,
 \end{equation}
which in the case $\lambda>0$ has a (negative) nonconstant scalar curvature given by
\begin{equation}
\mathcal{R}=-\lambda\frac{(N-1)(2N+3(N-2)\lambda \mathbf{x}^2)}{(1+\lambda \mathbf{x}^2)^3}.
\label{vcurv}
\end{equation}
As it was shown in~\cite{physicaD}, the Hamiltonian for the maximally superintegrable oscillator on this space is
\begin{equation}
H(\mathbf{x},\mathbf{p})=\frac{1}{2(1+\lambda \mathbf{x}^2)}\,(\mathbf{p}^2+\omega^2\mathbf{x}^2) \, ,
\label{eq}
\end{equation}
with real parameters $\lambda>0$ and $\omega\geq0$ and where $(\mathbf{x},\mathbf{p})$ are
canonically conjugated coordinates and momenta.
The curved space~\eqref{eq3}  is just the ND spherically symmetric generalization of the Darboux surface of
type III \cite{koen,Kal}.
Indeed, the Hamiltonian \eqref{eq} is a genuine $\lambda-$deformation of the well-known flat ND isotropic harmonic oscillator with frequency $\omega$, which is straightforwardly obtained in the limit $\lambda\to 0$:
\begin{equation}
\mathcal{H}_0=\frac{1}{2}(\mathbf{p}^2+\omega^2 \mathbf{x}^2), \qquad \mathrm{d}\mathbf{s}^2=\mathrm{d}\mathbf{x}^2, \qquad \mathcal{R}=0.
\end{equation}
Note that the generators of translations ({\em i.e.}~momenta operators) on the Darboux III space will not commute, since they are defined on a curved space. However, since this space has non-constant curvature the commutation rules between such translation operators will not be defined by a Lie algebra, opposedly to what happens in spaces with constant curvature (see, for instance, \cite{BHNK} and references therein).
It is also worth stressing that the Darboux III oscillator can be also interpreted as a nonlinear oscillator (yet defined on the ND Euclidean space) but whose mass $m$  is position-dependent in the form  $M(\mathbf{x})=1+\lambda\, \mathbf{x}^2$. The quantum version of the ND Darboux III oscillator is an exactly solvable quantum system that has been studied in detail in~\cite{PLA,cq3}.

A further type of symmetry-preserving generalization of the quantum version of the ND  isotropic oscillator is given by the so-called Dunkl oscillator, where usual derivatives are replaced by the Dunkl ones. The Dunkl operator introduced by Yang allows the definition of a type of deformed oscillator \cite{yang}, which includes of reflection symmetries \cite{wigner}, which allow the definition of new quantum integrable models \cite{integ,integ1,integ2}.
Several authors have generalized several integrable Hamiltonians by making use of the Dunkl operator, including Dunkl-Dirac \cite{dirac1,dirac2}, Dunkl-Klein-Gordon \cite{gordon1,gordon2}, Dunkl-pseudo harmonic oscillators \cite{pse}, and Dunkl-Duffin-Kemmer-Petiau oscillators \cite{dkp}. Recently, some of the authors have studied the symmetries defining the Schwinger-Dunkl algebra involving the Dunkl derivatives related to the -1 orthogonal polynomials of the Bannai-Ito algebra \cite{ito1,ito2,ito3,ito4}. Moreover, in Refs. \cite{two, three,2d} the two and three-dimensional curved Dunkl oscillators (associated to the so-called Jordan-Schwinger-Dunkl algebra) were studied, and the superintegrability of all these models was demonstrated.

The aim of this paper is the introduction of the Dunkl-Darboux III oscillator Hamiltonian, in which the nonlinear deformation arising in the Darboux III oscillator is supplemented by the additional reflection symmetries associated to Dunkl operators. By making use of all the abovementioned techniques, this new quantum model will be shown to be exactly solvable in the generic ND case, and its associated wavefunctions and energy eigenvalues will be explicitly presented. In particular, special attention will be payed to the $N=2$ case, since for that dimensionality we will show that a constant perpendicular magnetic field can be also introduced, thus giving rise to a biparametric family of exactly solvable models which provide the Dunkl-Darboux III generalization of the Landau levels. 

It is worth mentioning that the study of the two-dimensional Darboux III system under a constant magnetic field could be physically interesting since we recall that position-dependent mass functions of the type $M(x) = 1 + \lambda x^2$ are relevant in semiconductor heterostructures, for which their scattering properties \cite{dar1} as well as wave-packet revivals of the associated Schr\"odinger equation \cite{dar2} have been studied (also, see \cite{PDMsusy} and references therein for finite gap Hamiltonians with position-dependent mass and their connection with supersymmetry). On the other hand, Shannon information entropy for the eigenstates of the Darboux III oscillator in arbitrary dimensions has been recently analysed \cite{dar3}, and provides an excellent benchmark for the study the interplay between entropy and non-constant curvature (or position-dependent mass).

This paper is organized as follows. In Section 2, the Darboux III oscillator is reviewed in order to provide the necessary background needed for the rest of the paper. Also, the new quantum Hamiltonian for the 2D Darboux III oscillator in the presence of a constant magnetic field is introduced and fully solved. In particular, the modifications introduced in the Landau levels by the nonlinear deformation $\lambda$ and by the magnetic field $B$ ara analysed in detail. The well-known ND Dunkl oscillator is revisited in Section 3, and the new quantum model defining the 2D Dunkl oscillator in the presence of a magnetic field is also introduced and completely solved. Finally in Section 4 the ND Dunkl-Darboux III oscillator is defined and solved by making use of all the algebraic machinery presented in the two previous sections. Again, for the 2D case a constant magnetic field can be also included in the model, where  it is worth noticing that the Dunkl analogue of the angular momentum operator  allows for the separation of variables in terms of polar coordinates, and therefore leads to the explicit solution of the system. Finally, in Section 5, we give some concluding remarks.

\section{The Darboux III oscillator}

\subsection{The one-dimensional Darboux III oscillator}
In order to study the quantum systems associated to the one-dimensional version of the classical Hamiltonian \eqref{eq}, let us consider the standard
definitions for quantum position $\hat{x}$ and momentum $\hat{p}_x$ operators, with Lie
brackets and differential representation given by
\begin{equation}
\hat{x}\psi(x)=x\psi(x), \qquad \hat{p}_x\psi(x)=-i\hbar\frac{\partial\psi(x)}{\partial x}, \qquad [\hat{x},\hat{p}_x]=i\hbar.
\end{equation}
Hence, the well-known quantum version of the Hamiltonian
\eqref{eq} is given by~\cite{PLA,cq3}
\begin{equation}\label{e1}
\hat{\mathcal{H}}_{D'}=\frac{1}{(1+\lambda x^2)}\left[\frac{1}{2}\hat{p}_x^2+\frac{1}{2}\omega^2\hat{x}^2\right]=\frac{1}{(1+\lambda x^2)}\left[-\frac{\hbar^2}{2}\partial_x^2+\frac{1}{2}\omega^2x^2\right],
\end{equation}
which is formally self-adjoint provided the Hilbert space $L^{2}(\mathbb{R},\mathrm{d}x)$ is replaced by $L^{2}(\mathbb{R}^{2},(1+\lambda x^{2})\mathrm{d}x)$ where $(1+\lambda x^{2})$ is the weight function within the scalar product
\begin{equation}
	\langle \psi|\phi \rangle_\lambda=\int \overline{\psi(x)}\phi(x)(1+\lambda x^2)\mathrm{d}x.
\label{dhs}
\end{equation}
In the Hilbert space endowed with~\eqref{dhs},  the Hamiltonian  \eqref{e1} becomes Hermitian (we recall that real spectra can be also obtained from non-Hermitian Hamiltonians endowed with PT symmetry \cite{h1,h2,h3,h4,h5}).
The corresponding Schr\"{o}dinger equation $\hat{\mathcal{H}}_{D'}\psi(x)=E\psi(x)$ can be written as
\begin{equation}\label{eq2}
\frac{1}{2}\left[-\hbar^2\partial_x^2+\left(\omega^2-2\lambda E\right)x^2\right]\psi(x)=E\psi(x),
\end{equation}
which is just a quantum oscillator with an energy-dependent frequency given by
\begin{equation}
\Omega(E)=\sqrt{\omega^2-2\lambda E} \qquad\mathrm{whenever} \quad \omega^2>2\lambda E \, ,
\end{equation}and whose eigenvalues can be obtained by solving the equation 
\begin{equation}\label{eqE}
E=\hbar\Omega\left(n+\frac{1}{2}\right)\, .
\end{equation}
Therefore, we have an exactly solvable one-dimensional quantum model whose explicit eigenvalues come from solving the equation~\eqref{eqE}, and whose eigenfunctions $\psi_n(x)$ are given in analogy with the harmonic oscillator wavefunctions. In this way we obtain that \cite{PLA,cq3}
\begin{align}
& {E}_{n}=-\hbar^2\lambda\left(n+\frac{1}{2}\right)^2+\hbar\left(n+\frac{1}{2}\right)\sqrt{\lambda^2\hbar^2\left(n+\frac{1}{2}\right)^2+\omega^2}, \qquad n=0,1,2,\dots \label{esp1D}\\
& \psi_n(x)=A_{n}\left(\frac{\beta^2}{\pi}\right)^{\frac{1}{4}}\exp\left[-\beta^2x^2/2\right]H_{n}(\beta x), \qquad \beta=\sqrt{\frac{\Omega}{\hbar}},
\end{align}
where $H_n(\beta x)$ is the $n$-th Hermite polynomial and $A_n$ is a normalization constant.
Note that the real parameter $\lambda$ can be interpreted as a `deformation' parameter
governing the nonlinear behaviour of $\hat{\mathcal{H}}_{D'}$, and this parameter is deeply related to the
variable curvature of the underling Darboux space (see~\eqref{vcurv}). The case $\lambda =0$  leads to the well-known results for the harmonic oscillator Hamiltonian $\hat{\mathcal{H}}=-\frac{\hbar^2}{2}\partial_x^2+\frac{1}{2}\omega^2x^2$ with frequency $\omega$, which indeed does not depend on the energy $E$.


\subsection{The N-dimensional Darboux III oscillator}
In the ND generalization of the Darboux III system~\cite{physicaD,PLA,cq3}, the quantum position and momentum operators $(\hat{\mathbf{x}}, \hat{\mathbf{p}})$ are defined by
\begin{equation}
\hat{x}_i=x_i, \qquad \hat{p}_i=-i\hbar\frac{\partial}{\partial x_i}, \qquad [\hat{x}_i,\hat{p}_i]=i\hbar\delta_{ij} \, .
\end{equation}
Thus one can use the standard notation
\begin{equation}
\nabla=\left( \frac{\partial}{\partial x_1},\dots,\frac{\partial}{\partial x_N}\right), \qquad \Delta=\nabla^2=\frac{\partial^2}{\partial x_1^2}+\dots+\frac{\partial^2}{\partial x_N^2} \, ,
\end{equation}
and from the point of view of the so-called `Schr\"odinger' quantization~\cite{cq3}, the quantum Hamiltonian is given by
\begin{equation}\label{eq4}
\hat{H}_{D'}=\frac{1}{(1+\lambda \hat{\mathbf{x}}^2)}\left[\frac{1}{2}\hat{\mathbf{p}}^2+\frac{1}{2}\omega^2\hat{\mathbf{x}}^2\right]=\frac{1}{(1+\lambda \mathbf{x}^2)}\left[ -\frac{\hbar^2}{2}\nabla^2+\omega^2\mathbf{x}^2\right] \, .
\end{equation}
This quantization prescription preserves the maximal superintegrability of the system in a straightforward way due to the immediate quantum transcription of the $(2N-1)$ classical integrals of the motion, which are given by the operators
\begin{equation}\label{qintegrals}
  \hat I_i= \hat p_i^2- 2\lambda \, \hat q_i^2  {\hat H}_{D'}+ \omega^2 \hat q_i^2 ,\qquad i=1,\dots,N \, ,
\end{equation}
and therefore ${\hat H}_{D'}=\frac 12 \sum_{i=1}^N \hat I_i$.  

\begin{figure}[h!]
 	\begin{center}
 		\includegraphics[height=6.5cm,width=10cm]{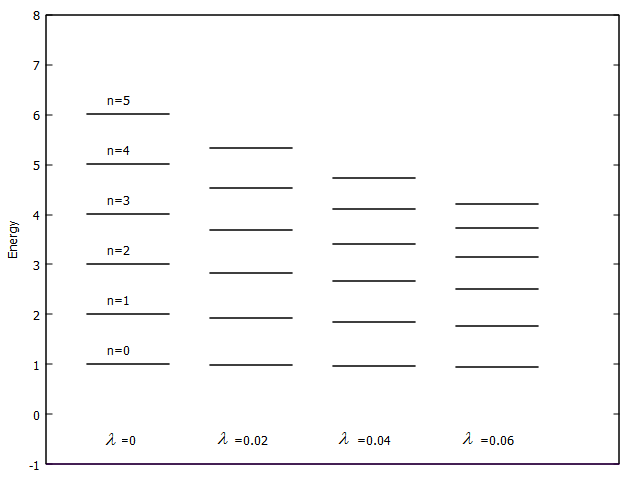}
 	\end{center}
 	\caption{Schematic representation of the energy levels \eqref{eq6} for the parameters $\hbar=\omega=1, N=2$ and $\lambda=\{0,0.02,0.04,0.04,0.06\}$ with quantum numbers $n=0,1,\dots,5$. When $\lambda=0$ the energies of the isotropic harmonic oscillator are recovered.}\label{fig0}
 \end{figure}

By taking into account a factorized wave function together with the eigenvalue equations for the quantum integrals $\hat I_i$
\begin{equation}\label{eq5}
\Psi (\mathbf{x})=\prod_{i=1}^{N}\psi_i(x_i), \qquad \tfrac 12 \,\hat I_i\,\Psi (\mathbf{x})=\nu_i\,\Psi (\mathbf{x}) \, ,
\end{equation} 
in which $\psi(x_i)\in L^2(\mathbb{R},\mathrm{d}x_i)$, the eigenvalues $\nu_i$ are shown to be~\cite{PLA,cq3} 
\begin{equation}
\nu_i\equiv\nu_i(E,n_i)=\hbar\Omega\left(n_i+\frac{1}{2}\right), \qquad n_i=0,1,2,\dots
\end{equation}
and the one-particle wave functions $\psi(x_i)$  are 
\begin{equation}
\psi_i(x_i)\equiv\psi_{n_i}(E,x_i)=A_{n_i}\left(\frac{\beta^2}{\pi}\right)^{\frac{1}{4}}\exp\left[-\beta^2x_i^2/2\right]H_{n_i}(\beta x_i) \, .
\end{equation}
From these expressions and by taking into account that ${\hat H}_{D'}=\frac 12 \sum_{i=1}^N \hat I_i$,  the discrete spectrum of the Hamiltonian \eqref{eq4} can be shown to be
 \begin{align}\label{eq6}
 \mathcal{E}_{n}=-\hbar^2\lambda\left(n+\frac{N}{2}\right)^2+\hbar\left(n+\frac{N}{2}\right)\sqrt{\lambda^2\hbar^2\left(n+\frac{N}{2}\right)^2+\omega^2}\, , 
 \qquad
 n=\sum_{i=1}^N n_i,\quad n=0,1,2\dots
 \end{align}
 Note that the spectrum of the Darboux III oscillator has the very same degeneracy as the ND quantum isotropic oscillator, but its eigenvalues are deformed in terms of the curvature parameter $\lambda$, thus giving rise to a nonlinear spacing among them. 
 This discrete spectrum for the first six levels of the $N=2$ model  is depicted in Fig.~\ref{fig0} for several values of $\lambda$. As we can observe, by increasing the deformation parameter $\lambda$, higher eigenvalues become more compressed, albeit the maximal degeneracy of the full spectrum is preserved with respect to the ND harmonic oscillator case.

\subsection{The 2D Darboux III oscillator  coupled to a constant magnetic field}

As it is well-known~\cite{Fock,Landau,Darwin,Ventriglia}, the isotropic oscillator on the flat 2D Euclidean space under the action of a constant perpendicular magnetic field in the $z$ direction is also an exactly solvable quantum model. In fact, if we consider the symmetric gauge
\begin{equation}\label{eq7}
\mathbf{A}=\left(-\frac{B\,y}{2},\frac{B\, x}{2}\right),
\end{equation}
the classical Hamiltonian reads
\begin{equation}\label{Lclass}
{\mathscr{H}}=\frac{1}{2}\left[\left(p_x-\frac{eA_x}{c}\right)^2+\left(p_y-\frac{eA_y}{c}\right)^2\right]+\frac{1}{2}\omega^2 (x^2+y^2),
\end{equation}
which can be written as
\begin{equation}
{\mathscr{H}}=\frac{1}{2}\left[(p_x^2+p_y^2)+\omega_c^2(x^2+y^2)-2\omega_c L_z\right]+\frac{1}{2}\omega^2 (x^2+y^2)\, ,
\end{equation} 
where $\omega_c=\frac{eB}{2c}$ is the Larmor frequency and $L_z=xp_y-yp_x$ is the angular momentum in the $z$ direction. Therefore, by introducing a modulation frequency given by
$\tilde{\omega}^2=\omega^2_c+\omega^2$, the Hamiltonian reads
\begin{equation}
{\mathscr{H}}=\frac{1}{2}\left[(p_x^2+p_y^2)+\tilde{\omega}^2(x^2+y^2)-2\omega_c L_z\right].
\end{equation}
The quantum counterpart of this Hamiltonian can be written as
\begin{equation}
\hat{\mathscr{H}}=\hat{\mathscr{H}}_{os}+ \hat{\mathscr{H}}_L =\frac{1}{2}\left[(\hat p_x^2+\hat p_y^2)+\tilde{\omega}^2(\hat x^2+ \hat y^2)]\right]-\omega_c \hat L_z \, ,
\end{equation}
and since both terms commute $[\hat{\mathscr{H}}_{os},\hat{\mathscr{H}}_L]=0$, the problem can be solved in terms of simultaneous eigenfunctions of both operators, where the first one is just the 2D isotropic oscillator with frequency $\tilde{\omega}$. Such eigenfunctions $\psi_{n,m}$ depend on two quantum numbers which can take values $n=0,1,2,\dots$ and $m=0.\pm 1,\pm 2,\dots$, and their eigenvalues $E_{n,m}$ are given by~\cite{Fock,Landau,Darwin,Ventriglia}
\begin{equation}\label{spect}
E_{n,m}=\hbar\,\tilde{\omega}\,(2n+|m|+1)  -\hbar \, \omega_c \, m.
\end{equation}

In the following we will show that if we consider that the initial oscillator has a position-dependent mass of the type $m(\mathbf{x})=1+\lambda \mathbf{x}^2=1+\lambda ({x}^2 + {y}^2)$, a new exactly solvable quantum model arises,  which can be called the 2D Darboux III oscillator on a constant magnetic field.

In the presence of an external uniform magnetic field ${B}$, this system would be defined as
\begin{equation}\label{eq8}
{\mathscr{H}}_{D'}=\frac{1}{2(1+\lambda \mathbf{x}^2)}\left[\left(p_x-\frac{eA_x}{c}\right)^2+\left(p_y-\frac{eA_y}{c}\right)^2+\frac{\omega^2}{2}(x^2+y^2)\right],
\end{equation}
where $e$ is the charge of the particle, and $\mathbf{A}$ is the magnetic vector potential in the symmetric gauge.
When substituting Eq. \eqref{eq7} into Eq. \eqref{eq8}, the Hamiltonian can be written as
\begin{equation}
{\mathscr{H}}_{D'}=\frac{1}{2(1+\lambda \mathbf{x}^2)}\left[(p_x^2+p_y^2)+\omega_c^2(x^2+y^2)-2\omega_c L_z+\frac{\omega^2}{2}(x^2+y^2)\right].
\end{equation} 
Therefore, with the same modulation frequency as above
$\tilde{\omega}^2=\omega^2_c+\omega^2$, the Hamiltonian reads
\begin{equation}
{\mathscr{H}}_{D'}=\frac{1}{2(1+\lambda \mathbf{x}^2)}\left[(p_x^2+p_y^2)+\tilde{\omega}^2(x^2+y^2)-2\omega_c L_z\right].
\end{equation}

In the following we show that the quantum version $\hat{\mathscr{H}}_{D'}$ of this model can be also exactly solved by following the same steps described in Section B for the Darboux oscillator. In particular, we will have that  $\hat{\mathscr{H}}_{D'}$, is an integrable $\lambda$-deformation of the quantum 2D harmonic oscillator Hamiltonian under the presence of a constant magnetic field.

If we use polar coordinates (see Apendix \ref{sec a}), then the Laplacian operator takes the form 
\begin{equation}
\nabla^2=\partial_x^2+\partial_y^2=\partial_r^2+\frac{1}{r}\partial_r+\frac{1}{r^2}L_z^2,
\end{equation} 
and under the usual quantization prescription we have $L_z=-i\hbar \partial_\theta$. Therefore, the quantum Hamiltonian reads 
\begin{equation}\label{eq9}
\hat{\mathscr{H}}_{D'}=\frac{1}{2(1+\lambda r^2)}\left[-\hbar^2\left(\partial_r^2+\frac{1}{r}\partial_r+\frac{1}{r^2}\partial_\theta^2\right)+2i\hbar \omega_c\partial_\theta+\tilde{\omega}^2r^2\right] \, .
\end{equation} 
The wavefunction in polar coordinates  $\psi(r,\theta)$ may be factorized as
\begin{equation}
\psi(r,\theta)=\frac{1}{\sqrt{r}}u(r)e^{im\theta},
\end{equation}
with $m\in \mathbb{Z}$, and  then the eigenvalue equation for the radial function $u(r)$  arising from Eq. \eqref{eq9} becomes
\begin{equation}
\left[-\frac{\hbar^2}{2}\partial_r^2+\frac{\hbar^2}{2r^2}\left(m^2-\frac{1}{4}\right)+\frac{1}{2}\tilde{\Omega}^2r^2\right]u(r)=\mathscr{E}_{nm}u(r),
\end{equation} 
where 
$\tilde{\Omega}=\sqrt{\tilde{\omega}^2-2\,{\mathscr{E}}_{nm}\,\lambda}$, and we will denote $\tilde{\mathscr{E}}_{nm}=\mathscr{E}_{nm}+\hbar m \omega_c$. Thus, the eigenspectrum in terms of different quantum numbers $(n,m)$ is obtained by solving the equation
\begin{equation}
\tilde{\mathscr{E}}_{nm}=\hbar\tilde{\Omega}(2n+|m|+1),
\end{equation} 
which leads to the following explicit expression for the eigenvalues:
\begin{equation}\label{eq0}
\mathscr{E}_{nm}=-\hbar m \omega_c-\hbar^2\lambda(2n+|m|+1)^2+\hbar\sqrt{\left[\hbar\lambda(2n+|m|+1)^2+m\omega_c\right]^2+\tilde{\omega}^2(2n+|m|+1)^2-m^2\omega_c^2} \, ,
\end{equation} 
where the quantum numbers are again $n=0,1,2,\dots$ and $m=0,\pm 1,\pm 2,\dots$. 
Eq. \eqref{eq0} shows the eigenvalues of the $\lambda$-deformed 2D quantum oscillator in a constant magnetic field. When we examine these eigenvalues, we observe that in the absence of a magnetic field and when the deformation is zero, $E_{n,m}=E_{n',m'}$  in all cases where $2n+|m| = 2n'+|m'|$. Moreover, when there is no magnetic field $(\omega_c\to0)$ but the deformation $\lambda$ is non-zero, we find that $E_{n,m}^{\lambda\neq 0}<E_{n,m}^{\lambda=0}$, thus resulting in identical degenerate states. This means that for a given state, when $\lambda$ increases the related energy decreases. On the other hand, for all values of $\lambda$, when the magnetic field is non zero, the degeneracy will be removed and $E_{n,|m|}<E_{n,-|m|}$.
When the magnetic field vanishes, so that when $\omega_c\to0$, this expression leads to the eigenvalues obtained in \eqref{eq6} for the Darboux III Hamiltonian $\hat{H}_{D'}$ in two-dimensions. On the other hand, in the limit $\lambda\to 0$ of constant mass we recover the eigenvalues~\eqref{spect} of the 2D oscillator within a constant magnetic field. Therefore, this model can be interpreted as a two-parametric $(\omega_c,\lambda)$-deformation of the 2D isotropic oscillator.
The solutions of Eq. \eqref{eq9} can be written in terms of the corresponding normalized eigenfunctions, namely
\begin{equation}
\psi_{n,m}(r,\theta)=\left[\frac{\Gamma(n+|m|)}{\Gamma(n+|m|+1)\pi}\right]^{1/2}\tilde{\beta}^{\tfrac{1}{2}+|m|}r^{|m|} e^{-\frac{\tilde{\beta}^2}{2}r^2}L_n^{|m|}\left(\tilde{\beta}^2r^2\right)e^{im\theta}, \qquad \tilde{\beta}=\frac{\tilde{\Omega}}{\hbar},
\end{equation}
where $L_n^{|m|}(u)$ are generalised Laguerre polynomials.

\begin{figure}[t]
 	\begin{center}
 		\includegraphics[height=8cm,width=15cm]{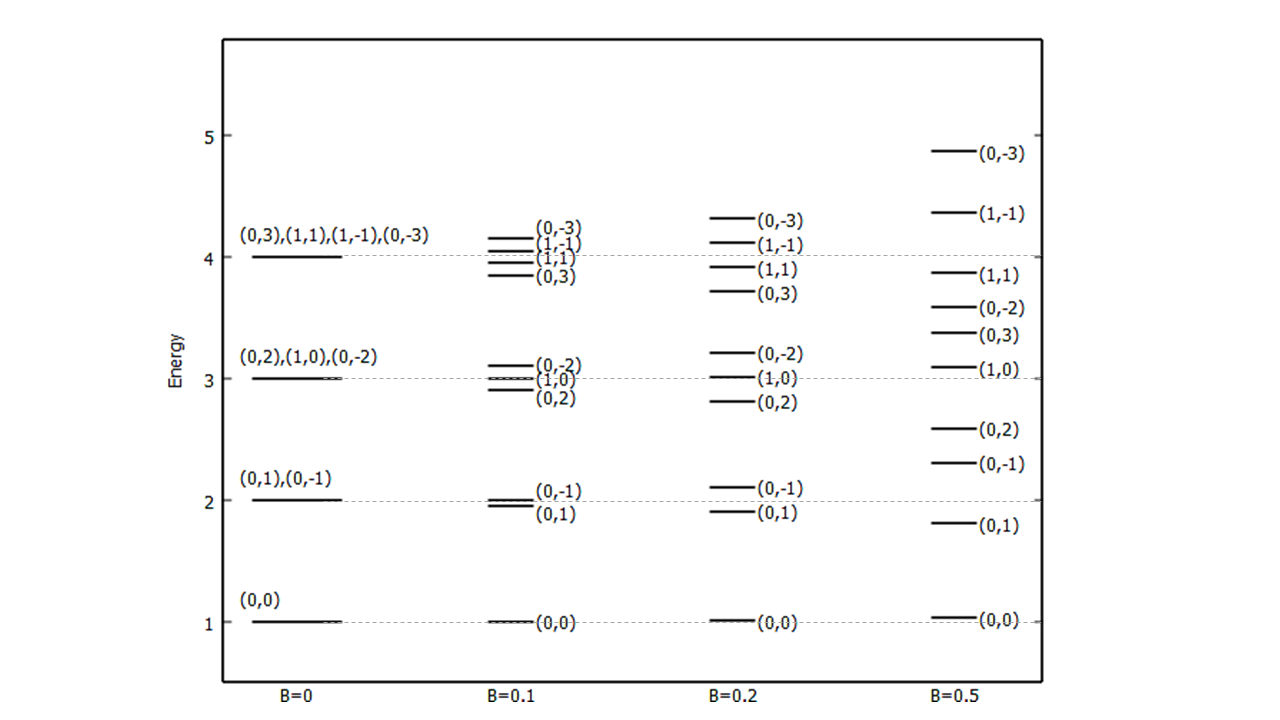}
 	\end{center}
 	\caption{The Landau levels are shown in terms of their quantum numbers $(n,m)$ for the usual case (null magnetic field)  and for $B=0.1, 0.2, 0.5$ (where some of the splittings induced by $m$ arise) by following \eqref{spect}, with $\hbar =\omega=1$. Note that here $\lambda= 0$.}\label{fig1}
 \end{figure}

\begin{figure}[h!]
  	\begin{center}
  		\includegraphics[height=8cm,width=15cm]{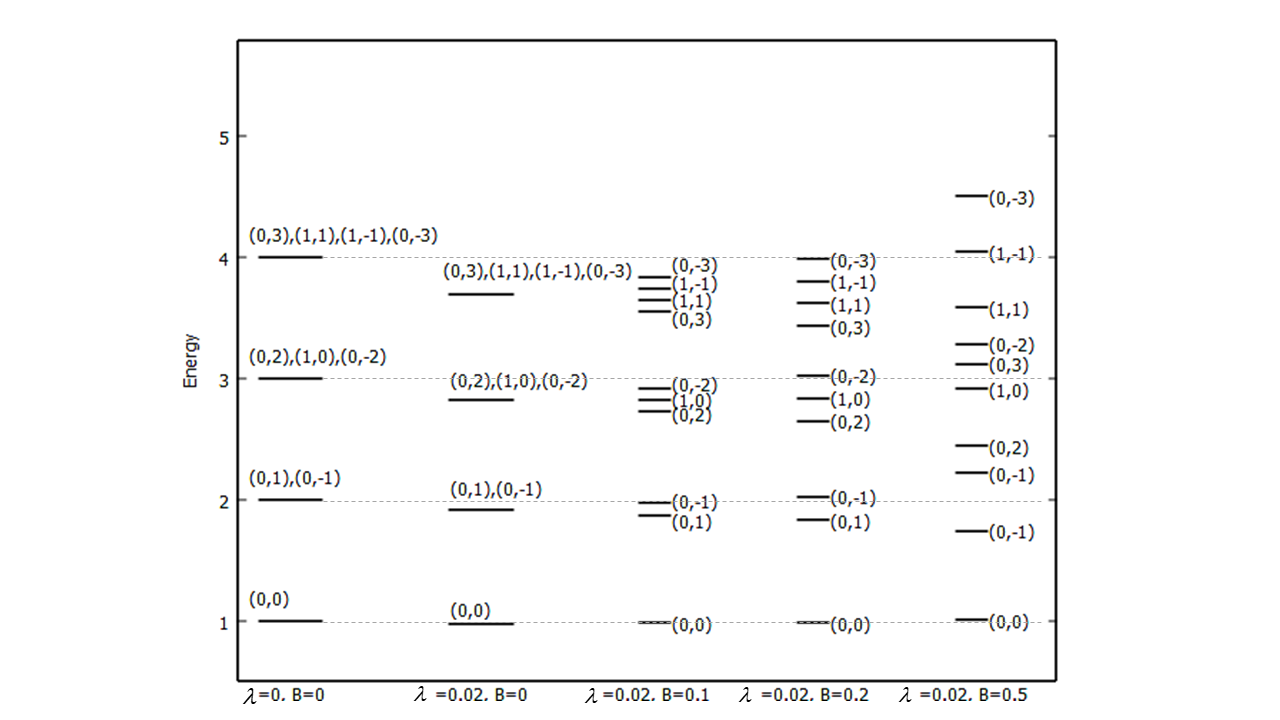}
  	\end{center}
  	\caption{Schematic representation of the Landau levels \eqref{eq0}  for the $\lambda-$deformed Darboux oscillator with quantum numbers $(n,m)$ for the $\lambda-$deformed Darboux oscillator, and also in conjunction with different magnetic fields of $B=0.1, 0.2, 0.5$. Again, we take $\hbar=\omega=1$. }\label{fig2}
  \end{figure}

The lowest energy levels of these exactly solvable quantum systems are plotted in Figs.~\ref{fig1} and \ref{fig2}. The first one presents the $\lambda=0$ case, {\em i.e.}, the Landau levels for a quantum oscillator under in a uniform magnetic field. The second one shows the influence of the $\lambda$-deformation
on such Landau levels for the cases with both null and non-vanishing magnetic field. Note that when $B=0$ there is a change of energies for the Landau levels, but they do not split in terms of $m$ since they correspond to the expression \eqref{eq6} for the Darboux II oscillator where $n\rightarrow 2n + |m|$ and $N\rightarrow 1$, and thus the level structure shown in Fig.~\ref{fig0} is reproduced. On the other hand, when $B\neq 0$ the combined effect of the magnetic field and the position-dependent mass governed by $\lambda$  can be appreciated.
 
\section{The Dunkl oscillator}

\subsection{The one-dimensional Dunkl oscillator}
We recall that the Dunkl oscillator is another exactly solvable quantum system constructed in terms of Dunkl operators. In the Dunkl
quantum mechanics the momentum operator is expressed in terms of the Dunkl derivative
instead of the ordinary derivative. These operators satisfy the following commutation rules \cite{ito1, defin,am,am1}
\begin{equation}\label{eq14}
\hat{P}_x=\frac{\hbar}{i}D_x^{\mu_x},\qquad  \hat{x}=x, \qquad [\hat{x}, \hat{P}_x]=i\hbar(1+2\mu_xR_x),
\end{equation}
where the Dunkl derivative is defined as
\begin{equation}
D_x^{\mu_x}=\partial_x+\frac{\mu_x}{x}(1-R_x),
\end{equation}
where $\mu_x$ is
real parameter such that $\mu_x>-1/2$, and $R_x$ denotes the reflection operator with respect to the plane $x=0$ which acts in the form $R_x f(x)=f(-x)$. The Hamiltonian for the one-dimensional Dunkl oscillator is defined by
\begin{equation}\label{eq16}
\hat{	\mathcal{H}}_{D}=-\frac{\hbar^2}{2} (D_x^{\mu_x})^2+\frac{1}{2}\omega^2 x^2.
\end{equation}
Since $[\mathcal{H}_{D},R_{x}]=0$, the eigenfunctions $f(x)$ may be chosen to have a well-defined parity  $R_x f(x)=e_x f(x)$ with $e_{x}=\pm1$.
We can also say that $q_{x}=0,1$ for $e_{x}=\pm1$ provided that $q_{x}=(1-e_{x})/2$. Hence, the eigenfunctions of the Dunkl oscillator are given by
\begin{equation}
	f_{n_{x}}^{e_x}(x)= e^{-\frac{\omega^2x^2}{2\hbar^2}} H_{n_{x}}^{\mu_{x}}(\tfrac{\omega}{\hbar} x), \qquad n_{x}=0,1,2,\dots
\end{equation}
where $H_{n_x}^{\mu_x}(\tfrac{\omega}{\hbar} x)$ are the generalized Hermite polynomials \cite{gener}
\begin{equation}
	H_{n_{x}}^{\mu_{x}}(\tfrac{\omega}{\hbar} x)=(-1)^n\sqrt{\frac{\omega n!}{\hbar\Gamma\left[n+\mu_{x}+q_{x}+\tfrac{1}{2}\right]}}x^{q_{x}} L_n^{\left(\mu_{x}+q_{x}-\tfrac{1}{2}\right)}(\tfrac{\omega^2}{\hbar^2}x^2), 
\end{equation} 
and the energies of the model are given by the eigenvalues
\begin{equation} \kappa_{n_{x}}^{e_x}\equiv\kappa(E',n_x,e_x)=\hbar\omega\left(n_{x}+\mu_{x}+\frac{1}{2}\right).
\end{equation}
We recall that the eigenfunctions of the one-dimensional Dunkl oscillator are normalized and
orthogonal on the weighted $L^2$ space endowed with the scalar product \cite{iner}
\begin{equation}
\langle g|f\rangle=\int \overline{g(x)} f(x) |x|^{2\mu_x} \mathrm{d}x.
\end{equation}


\subsection{The ND Dunkl oscillator}
The Dunkl oscillator can be generalized to $N$ dimensions \cite{wigner,ito1,ito2,ito3,ndimen}. In doing so, the Dunkl momentum operator $\hat{\mathbf{P}}$ associated with the reflection group $\mathbb{Z}^N_2$
is defined by
\begin{equation}
\hat{\mathbf{P}}=\left(\hat{P}_1,\dots,\hat{P}_N\right)=-i\hbar \left(D_{1}^{\mu_{x_1}}, D_{x_2}^{\mu_{2}},\dots,D_{x_N}^{\mu_{N}}\right). 
\end{equation}
In fact, we can write
\begin{align}
\nabla_{D}=\left( D_1^{\mu_1},\dots,D_N^{\mu_N}\right)=\sum_{i=1}^N
\partial_{x_i}+\frac{\mu_{i}}{x_i}(\mathbb{I}-\mathbf{R}_{i}),
\end{align}
where $\mathbb{I}$ is the identity operator, $\mu_i> -1/2$ are real numbers and the operator $\mathbf{R}_{i}$ is the reflection operator
with respect to the plane $x_i=0$, namely
\begin{equation}
\mathbf{R}_{i}f(x_i)=f(-x_i).
\end{equation}
The Dunkl-Laplacian operator for $\mathbb{Z}^N_2$ is therefore defined by
\begin{align}
\Delta_{D}=\nabla^2_{D}=(D_1^{\mu_1})^2+\dots+(D_N^{\mu_N})^2=\sum_{i=1}^N\partial_{x_i}^2+2\frac{\mu_{i}}{{x_i}}\partial_{x_i}-\frac{\mu_i}{{x_i}^2}(1-R_{i}).
\end{align}
This is a second-order differential-difference operator which reduces to the usual Laplacian $\nabla=\partial_{x_1}^2+\partial_{x_2}^2+\dots+\partial_{x_N}^2$, when $\mu_1=\mu_2=\dots=\mu_N=0$.
The Hamiltonian of the corresponding $N$-dimensional Dunkl oscillator problem is given by
\begin{equation}
\hat{H}_{D}=-\frac{\hbar^2}{2} \nabla^2_{D}+\frac{1}{2}\omega^2 \mathbf{x}^2 \, ,
\end{equation}
and since this system is separable, the related eigenfunctions and energy spectrum are given in the form
\begin{align}
& F_{n_{x_1},n_{x_2},\dots,n_{x_N}}^{e_1,e_2,\dots,e_N}(\mathbf{x})=\prod_{i=1}^{N}f_{n_{x_i}}^{e_i}(x_i),\\
& \bar{\mathcal{E}}_{n_{x_1},n_{x_2},\dots,n_{x_N}}^{e_1,e_2,\dots,e_N}=\sum_{i=1}^N \nu_{n_{x_i}}^{e_i}=\hbar \omega\left(\mathbf{n_x}+\mathbf{\mu_x}+\frac{N}{2}\right), 
\end{align}
in which $\mathbf{n_x}=\sum_{i=1}^{N}n_{x_i}$ and $\mathbf{\mu_x}=\sum_{i=1}^{N}\mu_i$.


\subsection{The 2D Dunkl oscillator coupled to a constant magnetic field }\label{sec3c}
In this section we will describe the 2D Dunkl oscillator in a constant magnetic field aligned along the z-axis, which will give rise to the Dunkl generalization of the Landau levels. The model Hamiltonian will be defined as
 \begin{align}\label{eq10}
\hat{ \mathscr{H}}_{D}=-\frac{\hbar^2}{2}\nabla_D^2+i\hbar\omega_c\left(xD_y^{\mu_y}-yD_x^{\mu_x}\right)+\frac{\tilde{\omega}^2}{2}(x^2+y^2),
 \end{align}
where $\nabla_D^2=(D_x^{\mu_x})^2+(D_y^{\mu_y})^2$, and the Dunkl derivatives $D_{x_i}$ are substituted instead of the standard partial derivatives $\partial_{x_i}$. By making use of polar coordinates,  the Dunkl-Laplacian operator can be written as
 \begin{equation}\label{eq17}
 \nabla_D^2=\partial_r^2+\frac{1}{r}\left(1+2\mu_x+2\mu_y\right)\partial_r-\frac{1}{r^2}\left(\mathcal{J}^2-2\mu_x\mu_y(1-R_xR_y)\right),
 \end{equation} 
where the angular momentum operator is $ \mathcal{J}\equiv i\hbar(xD_y^{\mu_y}-yD_x^{\mu_x})$ so that $\mathcal{J}^2=H_\theta+2\mu_x\mu_y(1-R_xR_y)$,  and $H_\theta$ is given in Eq. \eqref{eqa2} of Appendix  \ref{sec a}. By substituting Eq. \eqref{eq17} and the definition of the operator $\mathcal{J}$ into the Hamiltonian \eqref{eq10}, the following operator arises in polar coordinates:
 \begin{equation}\label{eq12}
\hat{ \mathscr{H}}_{D}=-\frac{\hbar^2}{2}\left(\frac{\partial^2}{\partial r^2}+\frac{1}{r}\left(1+2\mu_x+2\mu_y\right)\frac{\partial}{\partial r}\right)+\frac{\hbar^2}{2r^2}\left(\mathcal{J}^2-2\mu_x\mu_y(1-R_xR_y)\right)+\frac{\tilde{\omega}^2}{2}r^2+\hbar\omega_c\mathcal{J}.
 \end{equation} 
 
In the following we present the solutions of the two-dimensional Dunkl oscillator in the presence of the magnetic field. Our results are based on the fact that the operators $R_x,R_y$ commute with the one-dimensional Hamiltonians $\hat{\mathscr{H}}_{x},\hat{\mathscr{H}}_{y}$, respectively. Since reflection symmetry is understood as an element of a finite group $\mathbb{Z}_2$, having two irreducible representations for modes with even and odd parity, then the full Hilbert space of the system is splitted into even and odd sectors. Therefore, the resulting eigenfunctions for the two-dimensional Dunkl oscillator can be chosen
in such a way that they have a definite parity, as it is the case by considering the following two possibilities: 

 \subsubsection{Case $R_x=R_y$}
If both sectors are of the same type, $e_x = e_y = 1$ or $e_x = e_y =- 1$, this leads to $\epsilon\equiv e_xe_y=1$, where $e_x$ and $e_y$ are the eigenvalues of $R_x$ and $R_y$, respectively. By using the
 results of Appendix \ref{sec a}, $\mathcal{J}F_+=\sigma_+ F_+$ where $F_+=X_{m'}^{++}(\theta)\pm i X_{m'}^{--}(\theta)$ and $\sigma_+=\pm2\sqrt{m'(m'+\mu_x+\mu_y)}, \ m'\in \mathbb{N}$, where $X_{m'}^{++}$ and $X_{m'}^{--}$ are explicitly given in Eqs. \eqref{ap1} and \eqref{ap2}.
 \par
 For $e_x=e_y=1$, we have $\psi^{++}(r,\theta)=R^{++}(r)F_+(\theta)$. In order to find the radial solutions, we set $R^{++}(r)$ in the following eigenvalue equation
 \begin{equation}
 \frac{1}{2}\left[-\hbar^2\left(\frac{\partial^2}{\partial r^2}+\frac{1}{r}(1+2\mu_x+2\mu_y)\frac{\partial}{\partial r}+\frac{\sigma_+^2}{r^2}\right)+\tilde{\omega}^2r^2\right]R^{++}(r)=\tilde{\mathscr{E}}^+R^{++}(r),
 \end{equation}
where 
 $\tilde{\mathscr{E}}^+=\mathscr{E}^+-\hbar\omega_c\sigma_+$.
 Therefore, the radial solutions for the coupled Hamiltonian \eqref{eq12} are
  \begin{equation}
  R^{++}_n(r)=r^{-\mu_x-\mu_y+\sqrt{(\mu_x+\mu_y)^2+\sigma_+^2}} e^{-\frac{\omega^2}{2\hbar^2}r^2} L_n^{\sqrt{(\mu_x+\mu_y)^2+\sigma_+^2}}(\tfrac{\omega^2}{\hbar^2} r^2),
  \end{equation}
 and the energy spectrum is given by the explicit expression
 \begin{equation}\label{eq11}
 \mathscr{E}^+_{nm'}=\hbar\tilde{\omega}\left(2n+\sqrt{(\mu_x+\mu_y)^2+\sigma_+^2}+1\right)+\hbar \omega_c\sigma_+.
 \end{equation}
In the same manner, for $e_x=e_y=-1$, the wave function is $ \psi^{--}(r,\theta)=R^{--}(r)F_+(\theta)$ with the same $\mathcal{J}F_+=\sigma_+ F_+$. Consequently, the radial eigensolutions of \eqref{eq12} are again given by
 \begin{equation}
 R^{--}_n(r)=r^{-\mu_x-\mu_y+\sqrt{(\mu_x+\mu_y)^2+\sigma_+^2}} e^{-\frac{\beta^2}{2\hbar^2}r^2} L_n^{\sqrt{(\mu_x+\mu_y)^2+\sigma_+^2}}(\tfrac{\omega^2}{\hbar^2} r^2),
 \end{equation}
and also the eigenvalues are the same as in \eqref{eq11}. 
 \subsubsection{Case $R_x=-R_y$}
Here, we shall consider $\epsilon\equiv e_xe_y=-1$, and there are two possibilities: either $e_x = -1, e_y = 1$ or $e_x = 1, = e_y = -1$. As the previous case, from Appendix \ref{sec a}, we have that  $\mathcal{J}F_-=\sigma_- F_-$, where $F_-=X_{m'}^{+-}(\theta)\mp i X_{m'}^{-+}(\theta)$  in which $X_{m'}^{+-}(\theta)$ and $X_{m'}^{-+}(\theta)$ are given by Eqs. \eqref{ap3} and \eqref{ap4} of
the Appendix \ref{sec a}, so that we have $\sigma_-=\pm2\sqrt{(m'+\mu_x)(m'+\mu_y)}, \ m'\in \{\tfrac{1}{2},\tfrac{3}{2},\dots\}$.
\par
For the case $e_x=-1$ and $e_y=1$, we
find that the Hamiltonian \eqref{eq12} leads to the equation
\begin{equation}\label{eq13}
\frac{1}{2}\left[-\hbar^2\left(\frac{\partial^2}{\partial r^2}+\frac{1}{r}(1+2\mu_x+2\mu_y)\frac{\partial}{\partial r}+\frac{(\sigma_-^2-4\mu_x\mu_y)}{r^2}\right)+\tilde{\omega}^2r^2\right]R^{+-}(r)=\tilde{\mathscr{E}}^-R^{+-}(r)
\end{equation}  
where we have defined $\tilde{\mathscr{E}}^-=\mathscr{E}^--\hbar\omega_c\sigma_-$. 
As in the previous case, Eq. \eqref{eq13}  has admissible solutions 
$\psi^{+-}(r,\theta)=R^{+-}(r)F_-(\theta)$  where the radial solutions $R^{+-}(r)$ are obtained by solving Eq. \eqref{eq13}, and read
 \begin{equation}
 R^{+-}_n(r)=r^{-\mu_x-\mu_y+\sqrt{(\mu_x-\mu_y)^2+\sigma_-^2}} e^{-\frac{\omega^2}{2\hbar^2}r^2} L_n^{\sqrt{(\mu_x-\mu_y)^2+\sigma_-^2}}(\tfrac{\omega^2}{\hbar^2} r^2)
 \end{equation} 
 together with their eigenvalues
 \begin{equation}\label{e60}
 \mathscr{E}^-_{nm'}=\hbar\tilde{\omega}\left(2n+\sqrt{(\mu_x-\mu_y)^2+\sigma_-^2}+1\right)+\hbar \omega_c\sigma_-.
 \end{equation}

\begin{figure}[t]
	\begin{center}
		\includegraphics[height=9cm,width=16cm]{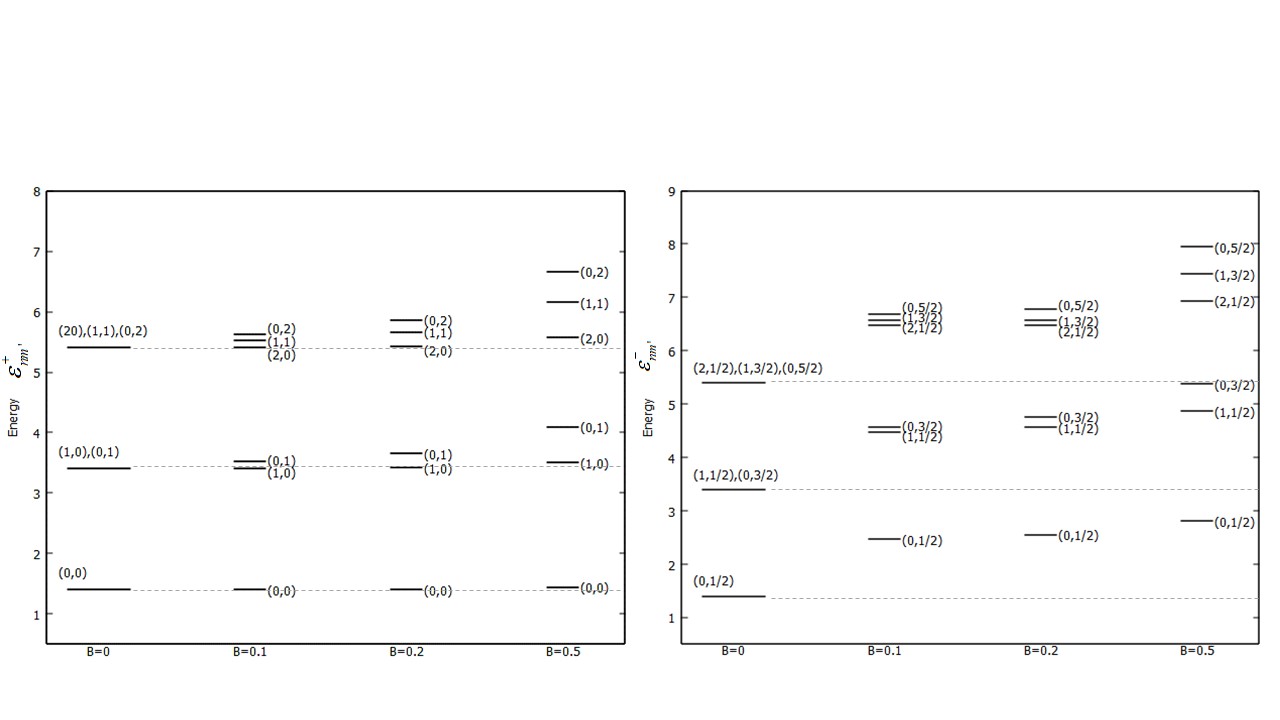}
	\end{center}
	\caption{Schematic representation of the first few Landau levels \eqref{eq11} and \eqref{e60} with quantum numbers $(n,m')$ for the null magnetic field (the usual Dunkl oscillator)  and for $B=0.1, 0.2, 0.5$. Note that here $\lambda=0$, $\hbar=\omega=1$ and the parameter values $\mu_x=\mu_y=0.02$. The left figure of the levels are those with the even sector $e_xe_y=+1$ and the right figure of the levels are those with the odd sector $e_xe_y=-1$. }\label{fig4}
\end{figure}

Similarly, for case $e_x=1$ and $e_y=-1$ from Eq. \eqref{eq13}, we obtain $R^{+-}(r)=R^{-+}(r)$ with the same eigenvalues $ \mathscr{E}^-_{nm'}$. The first few energy levels of the Dunkl oscillator systems are drawn in Fig. \ref{fig4}. It provides the Landau levels for the Dunkl oscillator under a uniform magnetic field, for which the deformed parameter is set as $\lambda=0$. These levels are the cases with both null and non-vanishing magnetic fields. Moreover, under the parity, $e_xe_y=\epsilon$, we present the numerical results to the energies given in \eqref{eq11} and \eqref{e60} in Fig. \ref{fig4}, which separates the spectrum of the system into two-level forms, even and odd. For both, it is seen the states with the same energy $N\rightarrow 2(n+|m'|)$, which these degeneracies can be broken to the different values in presence of the magnetic field.

\section{The Dunkl-Darboux III oscillator}

\subsection{The one-dimensional model}

At this point it seems natural to introduce a model called Dunkl-Darboux III oscillator $(\hat{\mathcal{H}}_{DD'})$ through the Hamiltonian
\begin{equation}\label{eq15}
\hat{\mathcal{H}}_{DD'}=\frac{1 }{(1+\lambda  x^2)}\left[-\frac{\hbar^2}{2}(D_x^{\mu_x})+\frac{\omega^2 }{2}x^2\right]=\frac{1 }{(1+\lambda x^2)} \hat{\mathcal{H}}_D \, ,
\end{equation}
which we will show to be a new exactly solvable quantum model.
This model describes a nonlinear oscillator with a momentum operator given in \eqref{eq14}, and can be interpreted as a new $\lambda-$deformation of the one-dimensional Dunkl oscillator. We recall that this deformation can be interpreted either as the motion on a space with nonconstant curvature or as the introduction of a position-dependent mass.
Note that when $\mu_x=0$, the Hamiltonian \eqref{eq15}
corresponds to the Darboux III quantum oscillator \eqref{e1},
and when $\lambda\rightarrow 0$, the Hamiltonian $\hat{\mathcal{H}}_{DD'}$ turns out to be the Dunkl oscillator given in \eqref{eq16}.
Indeed,  when both $\mu_x$ and $\lambda$ vanish, the Hamiltonian \eqref{eq15} is just the isotropic harmonic oscillator.

As a first relevant fact, it turns out that $\hat{\mathcal{H}}_{DD'}$ is formally self-adjoint on the Hilbert space $L^{2}(\mathbb{R}^{2}, \left(1+\lambda x^{2}\right) |x|^{2\mu_{x}})$ defined by the scalar product
\begin{equation}
\langle \varphi|\chi\rangle=\int \overline{\varphi(x)}\chi(x)(1+\lambda x^2)|x|^{2\mu_x}\mathrm{d}x.
\end{equation}
In the following we obtain the spectrum and eigenfunctions of the Dunkl-Darboux III quantum oscillator by solving  the Schr\"{o}dinger equation coming from the Hamiltonian \eqref{eq15}, namely
\begin{align}
\hat{\mathcal{H}}_{DD'} \, \varphi(x)&=\frac{1 }{(1+\lambda x^2)}\,\hat{\mathcal{H}}_D\,\varphi(x)=E^{\prime\prime}\varphi(x),\\
\hat{\mathcal{H}}_D\, \varphi(x)&=(1+\lambda x^2)\, E^{\prime\prime} \, \varphi(x).
\end{align}
Similarly to what we had for the Darboux III oscillator \eqref{eq2}, we are led to consider the following equation
\begin{equation}
\frac{1}{2}\left[-\hbar^2(D_x^{\mu_x})^2+(\omega^2-2E^{\prime\prime}\lambda)x^2\right] \varphi(x)=E^{\prime\prime} \varphi(x),
\end{equation}
and we can define the new energy-dependent frequency
\begin{equation}
\Omega'(E^{\prime\prime})=\sqrt{\omega^2-2E^{\prime\prime}\lambda}\quad \mathrm{where}\quad \omega^2>2E^{\prime\prime}\lambda.
\end{equation}
 Hence, eigenfunctions of $\hat{\mathcal{H}}_{DD'}$ can be expressed as
\begin{equation}\label{eq23}
f_{n_{x}}^{e_x}(x)= e^{-\frac{\beta'^2x^2}{2}} H_{n_{x}}^{\mu_{x}}(\beta' x), \qquad n_{x}=0,1,2,\dots
\end{equation}
in which
\begin{equation}
H_{n_{x}}^{\mu_{x}}(\beta' x)=(-1)^n\sqrt{\frac{\beta'^2 n!}{\Gamma\left[n+\mu_{x}+q_{x}+\tfrac{1}{2}\right]}}x^{q_{x}} L_n^{\left(\mu_{x}+q_{x}-\tfrac{1}{2}\right)}(\beta'^2x^2), \qquad \beta'=\sqrt{\frac{\Omega'}{\hbar}}.
\end{equation} 
And the energy spectrum is explicitly obtained by solving the equation
\begin{equation} \xi_{n_{x}}^{e_x}\equiv\xi(E'',n_x,e_x)=\hbar\Omega'\left(n_{x}+\mu_{x}+\frac{1}{2}\right).
\end{equation}
in terms of $E''$, which gives an expression identical to~\eqref{esp1D} with $n\rightarrow n_{x}+\mu_{x}$.


\subsection{The ND Dunkl-Darboux III oscillator}

This section is devoted to the N-dimensional generalization of the Dunkl-Darboux III oscillator Hamiltonian, which is defined as 
\begin{equation}\label{eq1}
\hat{\mathcal{H}}_{DD'}=\frac{1 }{(1+\lambda \mathbf{x}^2)}\left[-\frac{\hbar^2}{2}\Delta_D+\frac{\omega^2 }{2}\mathbf{x}^2\right]=\frac{1 }{(1+\lambda \mathbf{x}^2)}\, \hat{\mathcal{H}_D}\, .
\end{equation}
By following the same approach as in the previous subsection, the eigenvalue problem can be written as
\begin{equation} \varPhi_{n_{x_1},n_{x_2},\dots,n_{x_N},\lambda}^{e_1,e_2,\dots,e_N}(\mathbf{x})=\prod_{i=1}^{N}\varphi_{n_{x_i},\lambda}^{e_i}(x_i), \qquad (\hat{\mathcal{H}}_{DD'})_i\varPhi=\xi_{n_{x_i}}^{e_i}\varPhi \, .
\end{equation}
By taking into account the wave functions of Eq. \eqref{eq23}, eigenvalues turn out to be 
\begin{align}
\xi_{n_{x_i},\lambda}^{e_i}\equiv\xi_{n_{x_i},\lambda}^{e_i}(E'',n_{x_i})=\hbar \Omega'\left(\mathbf{n_{x_i}}+\mu_i+\frac{N}{2}\right),
\end{align} 
where once again $\Omega'(E'')=\sqrt{\omega^2-2E''\lambda}$. In this way we obtain  the following explicit expression for the energies of the ND Dunkl-Darboux III oscillator
$\sum_{i=1}^N \xi_{n_{x_i}}^{e_i}=\bar{\mathcal{E}}_{n_{x_1},n_{x_2},\dots,n_{x_N},\lambda}^{e_1,e_2,\dots,e_N}$
 in the form
 \begin{equation}
\bar{\mathcal{E}}_{n_{x_1},n_{x_2},\dots,n_{x_N},\lambda}^{e_1,e_2,\dots,e_N}=-\hbar^2\lambda\left(\mathbf{n_x}+\mathbf{\mu_x}+\frac{N}{2}\right)^2+\hbar\left(\mathbf{n_x}+\mathbf{\mu_x}+\frac{N}{2}\right)\sqrt{\lambda^2\hbar^2\left(\mathbf{n_x}+\mathbf{\mu_x}+\frac{N}{2}\right)^2+\omega^2}.
 \end{equation}
 \subsection{The Dunkl-Darboux III oscillator coupled to a constant magnetic field}
 
 In two dimensions, the Hamiltonian representing the Dunkl-Darboux III oscillator under a constant magnetic field aligned along the $z$-axis would be given by
 \begin{equation}
 \hat{\mathscr{H}}_{DD'}=\frac{1}{2(1+\lambda \mathbf{x}^2)}\left[-\hbar^2\nabla_D^2+2\hbar\omega_c\mathcal{J}+\tilde{\omega}^2(x^2+y^2)\right].
 \end{equation} 
 By making use of Eq. \eqref{eq17}, the radial Hamiltonian takes the form
\begin{equation}\label{eq18}
\hat{\mathscr{H}}_{DD'}=\frac{1}{(1+\lambda r^2)}\left[-\frac{\hbar^2}{2}\left(\frac{\partial^2}{\partial r^2}+\frac{1}{r}\left(1+2\mu_x+2\mu_y\right)\frac{\partial}{\partial r}\right)+\frac{\hbar^2}{2r^2}\left(\mathcal{J}^2-2\mu_x\mu_y(1-R_xR_y)\right)+\hbar\omega_c\mathcal{J}+\frac{\tilde{\omega}^2}{2}r^2\right].
\end{equation} 
It can be easily checked that $[R_x, \hat{\mathscr{H}}_{DD'}]=[R_y, \hat{\mathscr{H}}_{DD'}]=0$, and $[R_xR_y,\hat{\mathscr{H}}_{DD'}]=0$,  
and the operator $R_xR_y$ also commutes with the operator $\mathcal{J}$. Therefore we can make use of the results obtained in Sec. \ref{sec3c}. It what follows, we divide the radial solutions of Eq. \eqref{eq18} separately in two cases, namely $e_xe_y=\pm1$.
\subsubsection{Case $R_x=R_y$}
In this case, $e_xe_y=1$, by substituting the eigenvalue of the operator $\mathcal{J}$ equal to $\sigma_+$ in Eq. \eqref{eq18}, we can solve the radial Dunkl-Darboux III equation, which is
\begin{equation}
\frac{1}{2}\left[-\hbar^2\left(\frac{\partial^2}{\partial r^2}+\frac{1}{r}(1+2\mu_x+2\mu_y)\frac{\partial}{\partial r}+\frac{\sigma_+^2}{r^2}\right)+\tilde{\Omega}^2r^2\right]\mathscr{R}^{++}(r)=\tilde{\mathcal{E}}^+\mathscr{R}^{++}(r),
\end{equation} 
 where
$\tilde{\Omega}=\sqrt{\tilde{\omega}^2-2\lambda \mathcal{E}^+}$ and 
$\tilde{\mathcal{E}}^+=\mathcal{E}^+-\hbar\omega_L\sigma_+$. Hence, the energy spectrum of the Dunkl-Darboux III Hamiltonian given in \eqref{eq18} can be obtained by solving the equation
\begin{equation}\label{eq20}
\tilde{\mathcal{E}}^+_{\tilde{k}}=\hbar\tilde{\Omega}\left(2\tilde{k}+\sqrt{(\mu_x+\mu_y)^2+\sigma_+^2}+1\right).
\end{equation}
We immediately show that  the energy spectrum is
\begin{align}\label{e78}
\mathcal{E}^+_{\tilde{k}}&=\hbar \omega_c\sigma_+-\hbar^2\lambda \left(2\tilde{k}+\sqrt{(\mu_x+\mu_y)^2+\sigma_+^2}+1\right)^2\\
&+\hbar\sqrt{\left(\hbar\lambda \left(2\tilde{k}+\sqrt{(\mu_x+\mu_y)^2+\sigma_+^2}+1\right)^2-\omega_c\sigma_+\right)^2+\tilde{\omega}^2\left(2\tilde{k}+\sqrt{(\mu_x+\mu_y)^2+\sigma_+^2}+1\right)^2-\omega_c^2\sigma_+^2},\notag
\end{align}
with the eigenfunctions of the Dunkl-Darboux III oscillator given as
\begin{equation}\label{eq21}
\mathscr{R}^{++}_{\tilde{k}}(r)=r^{-\mu_x-\mu_y+\sqrt{(\mu_x+\mu_y)^2+\sigma_+^2}} \exp\left[-\frac{\beta^2}{2}r^2\right] L_{\tilde{k}}^{\sqrt{(\mu_x+\mu_y)^2+\sigma_+^2}}(\beta^2 r^2), \qquad \beta=\sqrt{\frac{\tilde{\Omega}}{\hbar}} \, .
\end{equation}
In order to obtain the explicit form
$\varphi^{--}_{\tilde{k}'}(r,\theta)=\mathscr{R}_{\tilde{k}'}^{--}(r)F_+(\theta)$ 
and its energy spectrum with two odd indexes, we should use Eqs. \eqref{eq20} and \eqref{eq21} with the different quantum number $\tilde{k}'$.

\subsubsection{Case $R_x=-R_y$}
Finally, we consider the second possibility given by $\epsilon=e_xe_y=-1$. Thus, from equation \eqref{eq18} we find that the radial part of the eigenfunctions
$
\psi^{+-}_{\tilde{\ell}}(r,\theta)=\mathscr{R}^{-+}_{\tilde{\ell}}(r)F_-(\theta)
$ are given by the equation
\begin{equation}
\frac{1}{2}\left[-\hbar^2\left(\frac{\partial^2}{\partial r^2}+\frac{1}{r}(1+2\mu_x+2\mu_y)\frac{\partial}{\partial r}+\frac{(\sigma_-^2-4\mu_x\mu_y)}{r^2}\right)+\tilde{\Omega}^2r^2\right]\mathscr{R}^{+-}_{\tilde{\ell}}(r)=\tilde{\mathcal{E}}^-_{\tilde{\ell}}\mathscr{R}^{+-}_{\tilde{\ell}}(r),
\end{equation} 
where now
$\tilde{\Omega}_{\tilde{\ell}}=\sqrt{\tilde{\omega}^2-2\lambda \mathcal{E}}^-_{\tilde{\ell}}$ and 
$\tilde{\mathcal{E}}^-_{\tilde{\ell}}=\mathcal{E}^-_{\tilde{\ell}}-\hbar\omega_L\sigma_-$.
Therefore, in this case the energy spectrum $\tilde{\mathcal{E}}^-_{\tilde{\ell}}$ is
\begin{equation}
\tilde{\mathcal{E}}^-_{\tilde{\ell}}=\hbar\tilde{\Omega}\left(2\tilde{\ell}+\sqrt{(\mu_x-\mu_y)^2+\sigma_-^2}+1\right),
\end{equation}
Substituting the expression for $\tilde{\Omega}_{\tilde{\ell}}$, the eigenvalues are found to be
\begin{align}\label{e82}
\mathcal{E}^-_{\tilde{\ell}}&=\hbar \omega_c\sigma_--\hbar^2\lambda \left(2\tilde{\ell}+\sqrt{(\mu_x-\mu_y)^2+\sigma_-^2}+1\right)^2\\
&+\hbar\sqrt{\left(\hbar\lambda \left(2\tilde{\ell}+\sqrt{(\mu_x-\mu_y)^2+\sigma_-^2}+1\right)^2-\omega_c\sigma_-\right)^2+\tilde{\omega}^2\left(2\tilde{\ell}+\sqrt{(\mu_x-\mu_y)^2+\sigma_-^2}+1\right)^2-\omega_c^2\sigma_-^2}\notag.
\end{align}
The explicit expression of eigenfunctions in terms of generalized Laguerre polynomials is
\begin{equation}\label{eq22}
\mathscr{R}^{+-}_{\tilde{\ell}}(r)=r^{-\mu_x-\mu_y+\sqrt{(\mu_x-\mu_y)^2+\sigma_-^2}} \exp\left[-\frac{\beta^2}{2}r^2\right] L_{\tilde{\ell}}^{\sqrt{(\mu_x-\mu_y)^2+\sigma_-^2}}(\beta^2 r^2).
\end{equation}
Under the condition  $e_x=-1$ and $e_y=1$, we find that the solutions of the eigenfunctions
$
\psi^{-+}_{\tilde{\ell}'}(r,\theta)=\mathscr{R}^{-+}_{\tilde{\ell}'}(r)F_-(\theta)
$ are the same given in Eq. \eqref{eq22} with the eigenvalues of $\mathcal{E}^-_{\tilde{\ell}}$ given in terms of the quantum number $\tilde{\ell}'$.
\begin{figure}[t]
	\begin{center}
		\includegraphics[height=9cm,width=16cm]{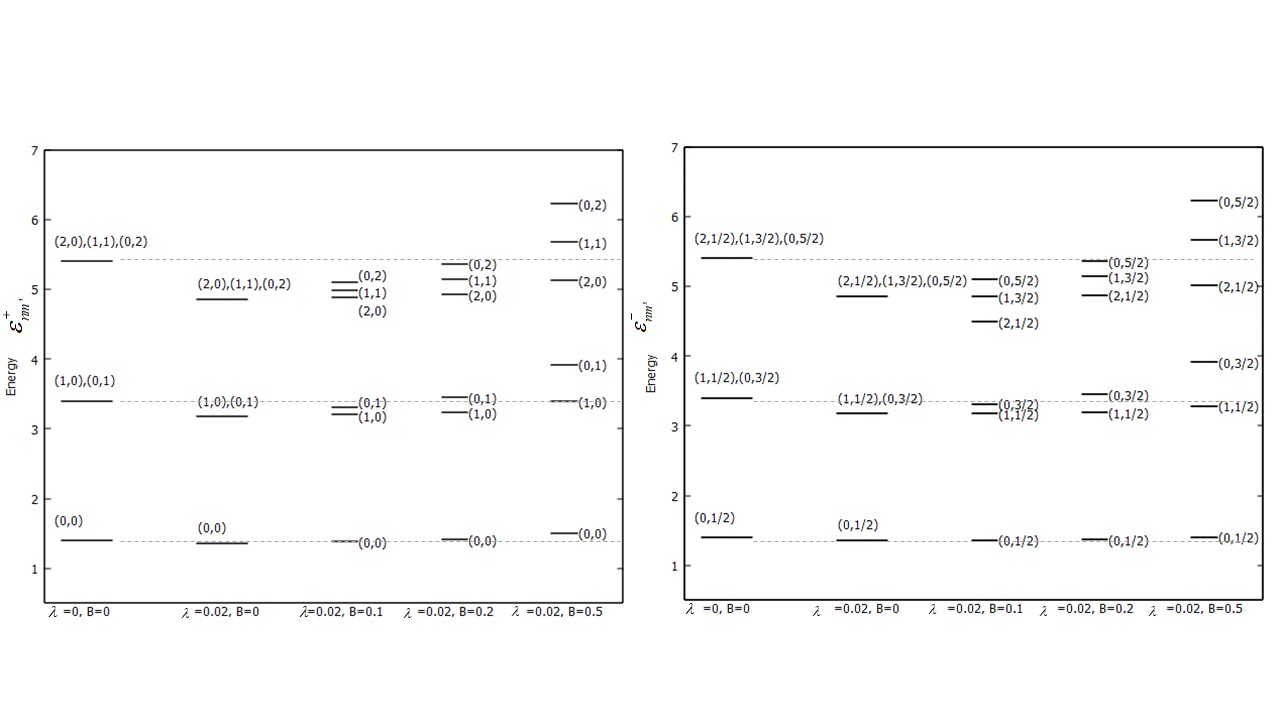}
	\end{center}
	\caption{Schematic representation of the first few Landau levels \eqref{e78} and \eqref{e82} with quantum numbers $(n,m')$ for the $\lambda-$deformed Dunkl-Darboux oscillator  with the different magnetic fields $B=0.1, 0.2, 0.5$, where the parameters are set with $\hbar=\omega=1$ and $\mu_x=\mu_y=0.02$. The left figure of the levels are those with the even sector $e_xe_y=+1$ and the right figure of the levels are those with the odd sector $e_xe_y=-1$.}\label{fig5}
\end{figure}
In Fig. \ref{fig5}, we have drawn numerical values for the first energy in the new model of the Dunkl-Darboux III, which is a combination of the two previous models. Plots of the Landau levels indicate the influence of the $\lambda-$deformation parameter for the Dunkl-Darboux III oscillator in the presence and absence of the magnetic field. 
It is enough for one of the parameters $\mu_i, B_i$, and $\lambda$ to go to zero, in this case, we will gain one of the modes discussed in the previous sections.

\section{Conclusions}

Summarizing, in this paper we have introduced a new familly of exactly solvable quantum models in $N$-dimensions that are constructed as integrable deformations of the isotropic oscillator Hamiltonian in two different, albeit compatible, directions. In the first one, a conformally flat space with nonconstant curvature is introduced as the background space for the model. On the other, the discrete symmetries provided by Dunkl operators can be also incorporated. The deformed symmetries allowed by both approaches turn out to be fully compatible, and provide all the tools needed in order to solve  the Dunkl-Darboux III model explicitly. Finally, in the 2D case a constant magnetic field can be also included by preserving the exact solvability of the Dunkl-Darboux III Hamiltonian. Moreover, the role played by the curvature parameter, the Dunkl eigenvalues and the magnetic field in the energies and degeneracies of the corresponding Landau levels can be explicitly shown.

\appendix

 \section{N-dimensional hyperspherical coordinates and Dunkl derivatives}\label{sec a}
 In analyzing the Darboux systems generalized with Dunkl derivatives on spaces of nonconstant curvature, the generalization to $N$-dimensional spaces of the hyperspherical coordinates $r, \theta_j$ is needed (see~\cite{cq2} for details). They are formed by a radial type coordinate $r=|\mathbf{x}|\in \mathbb{R}^+$ and $N-1$ angles $\theta_j$ such that $\theta_k\in [0,2\pi)$ for $k<N-1$ and $\theta_{N-1}\in [0,\pi)$. The coordinates are defined in correspondence with the Cartesian coordinates:
 \begin{equation}
 \hat{x}_j=\hat{r}\cos\hat{\theta}_j\prod_{k=1}^{j-1}\sin\hat{\theta}_k, \quad 1\leq j<N \qquad \hat{x}_N=\hat{r}\prod_{k=1}^{N-1}\sin\hat{\theta}_k,
 \end{equation}
 Hence the metric \eqref{eq3} takes the form
 \begin{equation}
 \mathrm{d}s^2=(1+\lambda r^2)(\mathrm{d}r^2+r^2\mathrm{d}\Omega^2), \qquad \mathrm{d}\Omega^2=\sum_{j=1}^{N-1}\mathrm{d}\theta_j^2\prod_{k=1}^{j-1}\sin^2\theta_k.
 \end{equation}
 Now, by considering the introduction of the quantum operators $p_r=-i\partial_r, p_{\theta_j}=-i\partial_{\theta_j}$, the corresponding Dunkl derivatives are defined as
 \begin{align}\label{eqa1}
 D_j&=\prod_{k=1}^{j-1}\sin\hat{\theta}_k\cos\hat{\theta}_j\partial_r+\frac{\cos\hat{\theta}_j}{\hat{r}}\sum_{i=1}^{j-1}\frac{\prod_{k=l+1}^{j-1}\sin\hat{\theta}_k}{\prod_{m=1}^{l-1}\sin\hat{\theta}_m}\cos\hat{\theta}_l\partial_{\theta_l}-\frac{\sin\hat{\theta}_j}{\hat{r}\prod_{k=1}^{j-1}\sin\hat{\theta}_k}\partial_{\theta_j}+\frac{\mu_j}{\hat{r}\cos\hat{\theta}_j\prod_{k=1}^{j-1}\sin\hat{\theta}_k}(1-R_j),\\
 D_N&=\prod_{k=1}^{N-1}\sin\hat{\theta}_k\partial_r+\frac{1}{\hat{r}}\sum_{l=1}^{N-1}\frac{\prod_{k=l+1}^{N-1}\sin\hat{\theta}_k}{\prod_{m=1}^{l-1}\sin\hat{\theta}_m}\cos\hat{\theta}_l\partial_{\theta_l}+\frac{\mu_N}{\hat{r}\prod_{k=1}^{N-1}\sin\hat{\theta}_k}(1-R_N).
 \end{align} 
 Then the explicit expression for the N-dimensional generalization of the Dunkl operators \eqref{eqa1} allows to write the Dunkl-Laplacian operator as
 \begin{equation}
 \Delta_D=\frac{\partial^2}{\partial r^2}+\frac{2\mu_j+N-1}{r}\frac{\partial}{\partial r}+\frac{1}{r^2}H_\theta,
 \end{equation}
where the operator $H_\theta$ is given in the two-dimensional case by
 \begin{align}\label{eqa2}
 H_\theta=-\frac{\partial^2}{\partial \theta^2}+\left(\mu_x\tan-\mu_y\cot\theta\right)\frac{\partial}{\partial \theta}+\frac{\mu_x(1-R_x)}{\cos^2\theta}+\frac{\mu_y(1-R_y)}{\sin^2\theta},
 \end{align}
 so that $H_\theta$ is related to the square of the Dunkl operator $\mathcal{J}$ in the following way 
 \begin{equation}
 \mathcal{J}^2=H_\theta+2\mu_x\mu_y(1-R_xR_y),
 \end{equation}
 where in polar coordinates the operator $\mathcal{J}$ takes the form
\begin{equation}
\mathcal{J}=i\left(\partial_\theta+\mu_y\cot\theta(1-R_y)-\mu_x\tan\theta(1-R_x)\right).
\end{equation} 
From the results presented in Refs. \cite{dirac2,pse,ito1}, we summarize the results for the eigenvalues of the angular momentum. In this way, we have the eigenvalue equation in the following form
 \begin{equation}\label{eqa3}
 \mathcal{J}F_\epsilon=\sigma_\epsilon F_\epsilon, 
 \end{equation}
 in which $\epsilon\equiv e_xe_y=\pm1$, and $e_x, e_y$ are the eigenvalues of the reflection operators $R_x$ and $R_y$, respectively. Since $\epsilon=\pm1$, eigenvalues and eigenfunctions of $\mathcal{J}$ can be obtained for $R_x=\epsilon R_y$. When $R_x=R_y$, that is $\epsilon=1$, the solutions of Eq. \eqref{eqa3} of $\mathcal{J}$ are given by
 \begin{equation}
 F_+=X^{++}_{m'}(\theta)\pm i X^{--}_{m'}(\theta)
 \end{equation}
with
  \begin{align}
  X^{++}_{m'}(\theta)&=\sqrt{\frac{(2m'+\mu_x+\mu_y)\Gamma(m'+\mu_x+\mu_y)m'!}{2\Gamma(m'+\mu_x+1/2)\Gamma(m'+\mu_y+1/2)}}P^{(\mu_x-1/2,\mu_y-1/2)}_{m'}(x),\label{ap1}\\
  X^{--}_{m'}(\theta)&=\sqrt{\frac{(2m'+\mu_x+\mu_y)\Gamma(m'+\mu_x+\mu_y)(m'-1)!}{2\Gamma(m'+\mu_x+1/2)\Gamma(m'+\mu_y+1/2)}}\sin\theta\cos\theta P^{(\mu_x+1/2,\mu_y+1/2)}_{m'-1}(x),\label{ap2}
  \end{align}
  where $m'$ is a non-negative integer, $\sigma_+=\pm2\sqrt{m'(m'+\mu_x+\mu_y)}$, and $P^{(\alpha,\beta)}_{m'}(x)$ are the Jacobi polynomials with $x=-\cos2\theta$. While for $R_x=-R_y (\epsilon=-1)$, the solutions of $\mathcal{J}$ leads to the expression 
  \begin{equation}
  F_-=X^{-+}_{m'}(\theta)\mp i X^{+-}_{m'}(\theta)
  \end{equation}
  with
  \begin{align}
  X^{+-}_{m'}(\theta)&=\sqrt{\frac{(2m'+\mu_x+\mu_y)\Gamma(m'+\mu_x+\mu_y+1/2)(m'-1/2)!}{2\Gamma(m'+\mu_x)\Gamma(m'+\mu_y+1)}}\sin\theta P^{(\mu_x-1/2,\mu_y+1/2)}_{m'-1/2}(x),\label{ap3}\\
  X^{-+}_{m'}(\theta)&=\sqrt{\frac{(2m'+\mu_x+\mu_y)\Gamma(m'+\mu_x+\mu_y+1/2)(m'-1/2)!}{2\Gamma(m'+\mu_x+1)\Gamma(m'+\mu_y)}}\cos\theta P^{(\mu_x+1/2,\mu_y-1/2)}_{m'-1/2}(x).\label{ap4}
  \end{align}
   where $m'$ is a positive half-integer, and $\sigma_-=\pm2\sqrt{(m'+\mu_x)(m'+\mu_y)}$.

\section*{Acknowledgements}
	The authors thank the referees for thoroughly reading our manuscript and for constructive suggestions that improved the original version of this paper. A. B. has been partially supported by Agencia Estatal de Investigaci\'on (Spain)  under grant  PID2019-106802GB-I00/AEI/10.13039/501100011033, by the Q-CAYLE Project funded by the Regional Government of Castilla y Le\'on (Junta de Castilla y Le\'on) and by the Ministry of Science and Innovation MICIN through the European Union funds NextGenerationEU (PRTR C17.I1). SH. D. acknowledges support from grant 20220355-SIP-IPN, Mexico. A. N. would like to thank the members of the Mathematical Physics Research Group of the University of Burgos for their kind assistance and hospitality.


\vskip-0.7cm



\begin{thebibliography}{99}



\bibitem{FMSUW65}
J. Fris, V. Mandrosov, Y.A. Smorodinsky, M. Uhlir, P. Winternitz,
{Phys. Lett.} {16} (1965) 354.

\bibitem{Ev90b}
N.W. Evans,
{Phys. Lett. A} {147} (1990) 483.

 \bibitem{Ev91}
 N.W. Evans, 
{J. Math. Phys.}  {32} (1991)  3369. 

 \bibitem{10}
C. Grosche, G.S. Pogosyan, A.N.   Sissakian,
{Fortschr. Phys.} {43}  (1995) 453.

 \bibitem{20}
 E.G. Kalnins, W. Miller, G.S. Pogosyan,
{J. Math. Phys.} {38}  (1997) 5416.

\bibitem{RS}
  M.F. Ra\~nada, M.  Santander, 
{J. Math. Phys.} {40}   (1999) 5026.

\bibitem{21}
E.G. Kalnins,  W. Miller, G.S. Pogosyan, 
{J. Phys. A: Math. Gen.} {33} (2000) 6791. 

\bibitem{BH07}
A. Ballesteros, F.J.  Herranz, 
{J. Phys. A:  Math. Theor.}
{40}  (2007) F51.   



\bibitem{physicaD}
A. Ballesteros, A. Enciso, F.J. Herranz, O. Ragnisco, {Physica D}   {237} (2008) 505.

\bibitem{PLA}
A. Ballesteros, A. Enciso, F.J. Herranz, O. Ragnisco, {Phys. Lett. A} {375} (2011) 1431.

\bibitem{cq0}
A. Ballesteros, A. Enciso, F.J. Herranz, O. Ragnisco, {Class. Quant. Grav.} {25} (2008) 165005.

\bibitem{cq1}
A. Ballesteros, A. Enciso, F.J. Herranz, O. Ragnisco, {Comm. Math. Phys.} {290} (2009) 1033.

\bibitem{cq2}
A. Ballesteros, A. Enciso, F.J. Herranz, O. Ragnisco, {Ann. Phys.} {324} (2009) 1219.

\bibitem{cq3}
A. Ballesteros, A. Enciso, F.J. Herranz, O. Ragnisco, D. Riglioni, {Ann. Phys.} {326} (2011) 2053.

\bibitem{ref15}
A. Ballesteros, A. Enciso, F.J. Herranz, O. Ragnisco, D. Riglioni, {Symm. Integrability  Geom.: Meth. Appl.} {7} (2011) 048.

\bibitem{latiniJPCS}
D. Latini, O. Ragnisco, A. Ballesteros, A. Enciso, F.J. Herranz, D. Riglioni, {J. Phys.: Conf. Ser.} {670} (2016) 012031.

\bibitem{bert1}
A. Ballesteros, A. Enciso, F.J. Herranz, O. Ragnisco, D. Riglioni, {J. Phys.: Conf. Ser.} {284} (2011) 012011.

\bibitem{cq6}
A. Ballesteros, A. Enciso, F.J. Herranz, O. Ragnisco, D. Riglioni, {J. Phys.: Conf. Ser.} {597} (2015) 012014.

\bibitem{cq7}
A. Ballesteros, I. Guti\'{e}rrez-Sagredo, P. Naranjo, {Phys. Lett. A} {381} (2017) 701.

\bibitem{koen}
G. Koenigs, in: G. Darboux (Ed.), Le\c{c}ons sur la th\'{e}orie g\'{e}n\'{e}rale des surfaces, vol. 4, Chelsea, New York, 1972, p. 368.

\bibitem{Kal}	
E.G. Kalnins, J.M. Kress, W.A. Miller Jr, P. Winternitz, {J. Math. Phys.} {44} (2003) 5811.	

\bibitem{BHNK}
A. Ballesteros, F.J. Herranz, J. Negro, S. Kuru, {Ann. Phys.} {373} (2016) 399.


\bibitem{yang}
L.M. Yang, {Phys. Rev.} {84} (1951) 788.
\bibitem{wigner}
C.F. Dunkl, {Trans. Amer. Math. Soc.} {311} (1989) 167.
\bibitem{integ}
T. Brzezi\'{n}ski, I.L. Egusquiza, A.J. Macfarlane, {Phys. Lett. B} {311} (1993) 202.
\bibitem{integ1}
K. Hikami, {J. Phys. Soc. Japan} {65} (1996) 394.
\bibitem{integ2}
S. Kakei, {J. Phys. A: Math. Gen.} {29} (1996) 619.
\bibitem{dirac1}
S. Sargolzaeipor, H. Hassanabadi, W.S. Chung, {Mod. Phys. Let. A} {33} (2018) 1850146.
\bibitem{dirac2}
R.D. Mota, D. Ojeda-Guill\'{e}n, M. Salazar-Ram\'{\i}rez, V.D. Granados, {Ann. Phys.} {411} (2019) 167964.
\bibitem{gordon1}
R.D. Mota, D. Ojeda-Guill\'{e}n, M. Salazar-Ram\'{\i}rez, V.D. Granados, {Mod. Phys. Let. A} {36} (2021) 2150171.
\bibitem{gordon2}
B. Hamil, B.C. L\"{u}tf\"{u}o\v{g}lu, arXiv preprint arXiv:2112.09948, (2021).
\bibitem{pse}
 R.D. Mota, D. Ojeda-Guill\'{e}n, M. Salazar-Ram\'{\i}rez, V.D. Granados, {Mod. Phys. Let. A} {36} (2021) 2150066.
\bibitem{dkp}
A. Merad, M. Merad, {Few-Body Sys.} {62} (2021) 1.
\bibitem{ito1}
V.X. Genest, M.E. Ismail, L. Vinet, A. Zhedanov, {J. Phys. A: Math. Theor.} {46} (2013) 145201.
\bibitem{ito2}
V.X. Genest, M.E. Ismail, L. Vinet, A. Zhedanov, {Comm. Math. Phys.} {329} (2014) 999. 
\bibitem{ito3}
V.X. Genest, L. Vinet, A. Zhedanov, {J. Phys.: Conf. Ser.} {512} (2014) 012010.
\bibitem{ito4}
V.X. Genest, A. Lapointe, L. Vinet, {Phys. Lett. A} {379} (2015) 923.
\bibitem{two}
A. Najafizade, H. Panahi, {Mod. Phys. Let. A} {37} (2022) 2250023.
\bibitem{three}
 S.H. Dong,  A. Najafizade, H. Panahi, W.S. Chung, H. Hassanabadi, arXiv preprint arXiv:2112.13546, (2021).
\bibitem{2d}
 P.S. Isaac, I. Marquette, {J. Phys. A: Math. Theor.} {49} (2016) 115201.

\bibitem{dar1}
R.~Ko\c c, M.~Koca, and G.~Sahinoglu, Eur. Phys. J. B-Cond. Matt. Comp. Sys. 48  (2005) 586.
\bibitem{dar2}
A.G. Schmidt, Phys. Lett. A, 353 (2006) 462.

\bibitem{PDMsusy}
R. Bravo, M.S. Plyushchay, {Phys. Rev. D} {93} (2016) 105023.

\bibitem{dar3}
 A. Ballesteros, I. Gutierrez-Sagredo, Phys. D: Nonlinear Phenomena, 445 (2023) 133618.
\bibitem{h1}
C.M. Bender, S. Boettcher, Phys. rev. lett. 80 (1998) 5243.
\bibitem{h2}
C. Yuce, Z. Oztas,  Sci. Rep. 8 (2018) 17416.
\bibitem{h3}
M. Znojil, F. Cannata, B. Bagchi, R. Roychoudhury,  Phys. Lett. B, 483 (2000) 284-289.
\bibitem{h4}
F.G. Scholtz, H.B. Geyer, F.J.W. Hahne, Ann. Phys. 213 (1992) 101.
\bibitem{h5}
F. Correa,  O. Lechtenfeld, J. High Energy Phys. 2019 (2019) 166.


\bibitem{Fock}
V. Fock, {Z. Phys.} {47} (1928) 446. 

\bibitem{Landau}
 L.D. Landau, {Z. Phys.} {64} (1930) 629. 

\bibitem{Darwin} 
G.C. Darwin, {Proc. Camb. Phil. Soc.} {27} (1931) 86.

\bibitem{Ventriglia}
 V.M. Ramaglia, B. Preziosi, A. Tagliacozzo, F. Ventriglia, In: E. Doni, R. Girlanda, G.P. Parravicini, A. Quattropani (Eds.), Progress in Electron Properties of Solids. Quantum Harmonic Oscillator in a Magnetic Field: An Example of Holomorphic Representation, vol. 10, Springer, Dordrecht, 1989, p. 451.
 
 \bibitem{defin}
V.X. Genest, L. Vinet, A. Zhedanov, {J. Phys. A: Math. Theor.} {46} (2013) 325201.
\bibitem{am}
F. Carrillo-Morales, F. Correa,  O. Lechtenfeld, J. High Energy Phys. 2021 (2021) 163.
\bibitem{am1}
F. Correa,  O. Lechtenfeld, J. High Energy Phys. 2017 (2017) 122.

\bibitem{gener}
M. Rosenblum, Generalized Hermite polynomials and the Bose-like oscillator calculus. In Nonselfadjoint operators and related topics, Birkh\"{a}user, Basel, 1994, p. 369.
\bibitem{iner}
 W.S. Chung, H. Hassanabadi, {Mod. Phys. Let. A} {34} (2019) 1950190.


\bibitem{ndimen}
S. Ghazouani, {Anal. Math. Phys.} {11} (2021) 35.



\end{thebibliography}
\end{document}